\documentclass[twocolumn,letterpaper,aps,prl,
superscriptaddress,showpacs,amsmath]{revtex4}
\usepackage{graphicx}
\newcommand{\vc}[1]{\boldsymbol{#1}}

\begin{document}
\title{Excitonic Magnetism in Van Vleck-type $d^4$ Mott Insulators}

\author{Giniyat Khaliullin}
\affiliation{Max Planck Institute for Solid State Research,
Heisenbergstrasse 1, D-70569 Stuttgart, Germany}

\begin{abstract}
In Mott insulators with the $t^4_{2g}$ electronic configuration such as of 
Re$^{3+}$, Ru$^{4+}$, Os$^{4+}$, Ir$^{5+}$ ions, spin-orbit coupling dictates 
a Van Vleck-type nonmagnetic ground state with angular momentum $J=0$, and 
the magnetic response is governed by gapped singlet-triplet excitations. We 
derive the exchange interactions between these excitons and study their 
collective behavior on different lattices. In perovskites, a conventional 
Bose condensation of excitons into a magnetic state is found, while an 
unexpected one-dimensional behavior supporting spin-liquid states emerges 
in honeycomb lattices, due to the bond-directional nature of exciton 
interactions in the case of 90$^\circ$ $d$-$p$-$d$ bonding geometry.   

\end{abstract}

\date{\today}

\maketitle

Many transition metal (TM) compounds fall into a category of Mott insulators 
where strong correlations suppress low-energy charge dynamics, but there 
remains rich physics due to unquenched spin and orbital magnetic moments that 
operate at energies below the charge (Mott) gap. Depending on spin-orbital 
structure of constituent ions and the nature of the chemical bonding of 
neighboring $d$ orbitals, the TM oxides host a great variety of magnetic 
phenomena~\cite{Ima98} ranging from classical orderings to quantum 
spin and orbital liquids. 

In broad terms, the magnetism of localized electrons in Mott insulators is 
governed by several factors: intraionic Hund's rules that form local spin 
$S$ and orbital $L$ moments; spin-orbit coupling (SOC) that tends to bind 
them into a total angular momentum $\vc{J}=\vc{S}+\vc{L}$; crystal fields 
which split $d$ levels and suppress the $L$ moment, acting thereby against SOC; 
and, finally, intersite superexchange (SE) interactions which establish 
a long-range coherence between spins and orbitals. 

In Mott insulators with $d$ orbitals of $e_g$ symmetry like manganites and
cuprates, $L$ moment is fully quenched in the ground state (GS), and one is 
left with spin-only magnetism. In contrast, TM ions with threefold $t_{2g}$ 
orbital degeneracy possess an effective orbital angular momentum $L=1$, and 
a complex interplay between unquenched SOC and SE interactions emerges. 

In TM oxides with odd number of electrons on the $d$ shell, $S$ and $J$ are 
half-integer; hence, the ionic GS is Kramers degenerate and magnetically 
active. The main effect of SOC in this case is to convert the original exchange 
interactions among $S$ and $L$ moments into an effective $J$-Hamiltonian 
operating within the lowest Kramers $J$-manifold. The $t_{2g}$ orbital  
$L$-interactions are bond dependent and highly frustrated~\cite{Kha02};
consequently, the $J$-Hamiltonians inherit this property, too. 
In short, SOC replaces $\vc S$ and $\vc L$ moments by $\vc J$ that obeys the 
same spin-commutation rules, but resulting magnetic states may obtain 
a nontrivial structure, as found for 
$d^5(J\!\!=\!\!1/2)$~\cite{Kha05,Che08,Jac09,Shi09,Kim09,Cha10,Bha12,Wit13} and 
$d^1(J\!\!=\!\!3/2)$~\cite{Jac09V,Che10} compounds. Similar SOC effects can be  
realized also in non-Kramers $d^2$ oxides~\cite{Per09,Che11} with $J=2$. 
 
A conceptually different situation can be encountered in Mott 
insulators with TM ions of Van Vleck-type, i.e. when SOC imposes nonmagnetic 
GS with $J=0$ and magnetism is entirely due to virtual transitions 
to higher levels with finite $J$. Such ''nonmagnetic'' Mott insulators are 
natural for $4d$ and $5d$ TM ions with $t^4_{2g}$ configuration, e.g., 
Re$^{3+}$, Ru$^{4+}$, Os$^{4+}$, and Ir$^{5+}$. These ions realize low-spin $S=1$
state because of moderate Hund's coupling $J_H$ (compared to $10Dq$ octahedral 
splitting), and, at the same time, SOC $\lambda(\vc{S}\cdot\vc{L})$ is strong 
enough to stabilize $J=0$ state gaining energy $\lambda$ relative to the 
excited $J=1$ triplet. Since the singlet-triplet splitting for these ions 
$\lambda\sim 50-200$~meV~\cite{Abr70,Fig00} is comparable to SE energy scales 
$4t^2/U \sim 50-100$~meV, we may expect magnetic condensation of Van Vleck
excitons. This brings us to the ''singlet-triplet'' physics widely discussed 
in the literature in various contexts: magnon condensation in quantum dimer 
models~\cite{Mat04,Gia08,Rue08,Sac11,Kul11}, bilayer magnets~\cite{Som01}, 
excitons in rare-earth filled-skutterudites~\cite{Kuw05,Ray08,Shi04}, a 
curious case of $e_g$ orbital FeSc$_2$S$_4$~\cite{Che09}, spin-state
transition in Fe-pnictides~\cite{Cha13}, etc. The underlying physics and, 
hence, the energy scales involved in the present case are of course different 
from the above examples.  

In this Letter, we develop a microscopic theory of magnetism for 
Van Vleck-type $d^4$ Mott insulators. First, we derive $S$ and $L$ based SE
Hamiltonian and map it onto a singlet-triplet low-energy Hilbert space.
We then show, taking perovskite lattices as an example, how an excitonic 
magnetic order, magnons and the amplitude (''Higgs'') modes do emerge in 
the model. Considering the model on a honeycomb lattice, we reveal the emergent 
one-dimensional dynamics of Van Vleck excitons, and discuss possible 
implications of this observation.  

{\it The spin-orbital superexchange}.-- Kugel-Khomskii type interactions 
between $t^4_{2g}$ ions are derived in a standard way, by integrating out 
oxygen-mediated $d$-$p$-$d$ electron hoppings. We label $d_{yz},d_{zx},d_{xy}$
orbitals by $a,b,c$, respectively. In 180$^\circ$ (90$^\circ$) 
$d$-$p$-$d$ bonding geometry corresponding to corner-shared (edge-shared)
octahedra, two orbitals are active on a given bond (see Fig.~2 of 
Ref.~\cite{Jac09}), while the third one, say, $\gamma$, is not; 
accordingly, this bond is denoted by $\gamma$. Then, nearest-neighbor
hoppings on the $c$ bond read as $t(a^{\dagger}_ia_j+b^{\dagger}_ib_j+h.c.)$
for 180$^\circ$ and $t(a^{\dagger}_ib_j+b^{\dagger}_ia_j+h.c.)$ for 90$^\circ$ 
geometries. In calculations, it is helpful to introduce $A,B,C$ operators 
that represent three different orbital configurations 
$A=\{a^2bc\},B=\{ab^2c\},C=\{abc^2\}$ of the $t^4_{2g}$ shell and 
its effective $L=1$ momentum $L_x=-i(B^{\dagger}C-C^{\dagger}B)$, etc.,  
similar to the $d^1$ case~\cite{Kha02}. [There is one-to-one correspondence
$(A,B,C)\leftrightarrow (a,b,c)$ and $L^\alpha \leftrightarrow l^\alpha$ between 
$d^4$ and $d^1$ orbital configurations.] 

The resulting spin-orbital Hamiltonian reads as 
$H=\frac{t^2}{U}\sum_{<ij>}[(\vc{S_i}\cdot\vc{S_j}+1)O_{ij}^{(\gamma)}+
(L_i^\gamma)^2+(L_j^\gamma)^2]$, where $U\gg t,\lambda,J_H$ is Hubbard 
repulsion. The bond-dependent orbital operator $O^{(\gamma)}$ depends on the 
above $A,B,C$; we show it directly in terms of $\vc{L}$: 
\begin{equation}
O_{ij}^{(c)}=(L_i^xL_j^x)^2+(L_i^yL_j^y)^2+L_i^xL_i^yL_j^yL_j^x+L_i^yL_i^xL_j^xL_j^y. 
\label{eq:LL}
\end{equation}
This result holds for 180$^\circ$ bonding. For 90$^\circ$ geometry, one has 
simply to interchange $L_j^x \leftrightarrow L_j^y$; this can be traced back 
to the $a_j\leftrightarrow b_j$ relation between 180$^\circ$ and 90$^\circ$ 
hoppings given above. Operators $O^{(a)}$ and $O^{(b)}$ for $a,b$ bonds follow
from cyclic permutations among $L_x,L_y,L_z$.    

The above model $H$ operates within $|M_S,M_L\rangle$ basis. We project it 
onto the low-energy subspace spanned by the GS $J=0$ singlet 
$|0\rangle=\frac{1}{\sqrt 3}(|1,-1\rangle-|0,0\rangle+|-1,1\rangle)$, 
and $J=1$ triplet 
$|T_0\rangle=\frac{1}{\sqrt 2}(|1,-1\rangle-|-1,1\rangle)$,
$|T_{\pm 1}\rangle=\pm\frac{1}{\sqrt 2}(|\pm 1,0\rangle-|0,\pm 1\rangle)$
at energy $E_T=\lambda$, as dictated by local SOC.   
(The high-energy $J=2$ level at $3\lambda$ is neglected.)  
Calculating matrix elements of $S^\alpha$, $L^\alpha$, and their
combinations within this Hilbert space, we represent them in terms of 
hard-core ''triplon'' $\vc T$ with the Cartesian components 
$T_x=\frac{1}{i\sqrt 2}(T_1-T_{-1})$, $T_y=\frac{1}{\sqrt 2}(T_1+T_{-1})$, 
$T_z=iT_0$, and ''spin'' $\vc J = -i(\vc T^\dagger\times\vc T)$. For instance, 
$\vc S=-i\sqrt\frac{2}{3} (\vc T- \vc T^\dagger)+ \frac{1}{2}\vc J$, 
$\vc L= i\sqrt\frac{2}{3} (\vc T- \vc T^\dagger)+ \frac{1}{2}\vc J$.  
A projection $H(S,L)\rightarrow H(T,J)$ results in the effective 
singlet-triplet models $H_{eff}(180^\circ)$ and $H_{eff}(90^\circ)$ 
discussed shortly below. 

In terms of $\vc T$ and $\vc J$, magnetic moment of a $t^4_{2g}$ shell 
$\vc M=2\vc S -\vc L$ reads
as $\vc M=-i\sqrt 6 \;(\vc T- \vc T^\dagger) + g_J\vc J$ with $g_J=1/2$, or 
$\vc M=2\sqrt 6 \;\vc v + g_J\vc J$, introducing real fields $\vc u$ and 
$\vc v$ as $\vc T=\vc u +i\vc v$ with $u^2+v^2\leq 1$~\cite{note_uv}. 
The two-component structure of $\vc M$ highlights physical distinction between 
conventional Mott insulators where $\vc M$ is simply $g_J\vc J$ with finite 
$J$ in the GS, and the present case where the magnetic moment resides 
predominantly on singlet-triplet Van Vleck transitions represented by 
$\vc T$ exciton (hence the term ''excitonic magnetism''). 
On formal side, these two components obey different commutation rules, 
hard-core boson $\vc T$ vs spin $\vc J$; consequently, magnetic order is 
realized here as Bose condensation of $\vc T$ particles, instead of the usual 
freezing of the preexisting $\vc J$ moments. The above equations for $\vc S$ 
and $\vc L$ make it also clear that $\vc T$ condensation implies a 
condensation of $\vc S$ and $\vc L$ moments resulting in finite $\vc M$, 
while the sum $\vc S +\vc L=\vc J$ may still fluctuate~\cite{note_J}. As in 
singlet-triplet models in general, the magnetic exciton condensation in 
$t^4_{2g}$ Van Vleck systems requires a critical exchange coupling $t^2/U$, 
so there will be magnetic order-disorder critical point that can be tuned by 
pressure, doping, etc. 

{\it Singlet-triplet model $H_{eff}(180^\circ)$}.-- This case applies to 
perovskites like {\it AB}O$_3$ or {\it A}$_2${\it B}O$_4$ with corner-shared 
{\it B}O$_6$ octahedra (e.g., Ca$_2$RuO$_4$). We shape the model in the form 
of $H_{eff}=\lambda\sum_in_i+\frac{t^2}{U}\sum_{<ij>} (h_2+h_3+h_4)_{ij}^{(\gamma)}$, 
where $h_2$ term is quadratic in $T$ bosons, while $h_3$ and $h_4$ represent
three- and four-boson interactions~\cite{note_ni}. For the $\gamma=c$ bond,
\begin{eqnarray}
\label{eq:180h2}
h^{(c)}_2\!\!&=&\!\frac{11}{3}\vc v_i\!\cdot\!\vc v_j - v_{iz}v_{jz} 
+\frac{1}{3}(\vc u_i\!\cdot\!\vc u_j- u_{iz}u_{jz}), \\ 
\label{eq:180h3} \nonumber
h^{(c)}_3\!\!&=&\!\frac{1}{\sqrt{24}}(\vc v_i\!\cdot\!\vc J_j \!+ \!v_{iz}J_{jz} 
+u_{ix}Q_{jx} \!- \!u_{iy}Q_{jy}) + (i\!\leftrightarrow\!j), \\  
\label{eq:180h4} \nonumber
h^{(c)}_4\!\!&=&\!\frac{3}{4}d^\dagger_{ij}d_{ij} 
+\frac{1}{2}\vc J_i\!\cdot\!\vc J_j 
+\frac{1}{4}(J_{iz}J_{jz}\!+\!J^2_{iz}J^2_{jz})-\frac{5}{36}n_in_j. 
\end{eqnarray}
$h^{(\gamma)}$ for $\gamma=a,b$ follow from permutations among $x,y,z$. 
$n=\sum_{\gamma}T^\dagger_{\gamma}T_{\gamma}$, while 
$Q_x=-(T^\dagger_yT_z+T^\dagger_zT_y)$, etc, are quadrupole operators of $T_{2g}$ 
symmetry~\cite{note_Q}. As expected, $h_4$ contains a biquadratic Heisenberg 
coupling; we show it here via bond-singlet operator 
$d_{ij}=\frac{1}{\sqrt3}(\vc T_i\!\cdot\!\vc T_j)$ using the identity 
$(\vc J_i\!\cdot\!\vc J_j)^2=3d^\dagger_{ij}d_{ij}+n_in_j$. 
 
We quantify exchange interaction by $\kappa=4t^2/U$. On a cubic (square)
lattice, the model undergoes a magnetic phase transition at 
$\kappa_c\simeq\frac{2}{5}\lambda$ ($\kappa_c\simeq\frac{3}{5}\lambda$), 
due to condensation of a dipolar $\vc v$ part of the $\vc T$ bosons. The density
of out-of-condensate $T$ particles and, hence, $\vc J$ and $Q_\alpha$ are very
small near critical $\kappa$, e.g., $\langle J_z^2\rangle_{\kappa_c} \sim 1/8z$
with $z=6 (4)$ for a cubic (square) lattice; thus, the interactions $h_{3,4}$ 
are not of a qualitative importance for the $180^\circ$ case, and we focus on a 
quadratic part $H_2$ of $H_{eff}$. Also, bond-dependent terms in $h_2$ are weak
and unessential in $180^\circ$ geometry, so we may average them out: 
$v_{i\gamma}v_{j\gamma}\rightarrow \vc v_i\!\cdot\!\vc v_j/3$ for 
simplicity~\cite{note_av}. The resulting hard-core boson Hamiltonian  
\begin{equation}
H_2= \lambda\sum_in_i+\kappa\;\frac{2}{9}
\sum_{ij} [\vc T^\dagger_i \!\cdot\!\vc T_j
-\frac{7}{16}(\vc T_i\!\cdot\!\vc T_j + H.c.)] 
\label{eq:180}
\end{equation}
is treated in a standard way familiar from ''singlet-triplet model'' 
literature (see, e.g., Refs.~\onlinecite{Mat04,Som01}). 

In a paramagnetic phase, $\kappa<\kappa_c$, magnetic excitations are
degenerate, and their dispersion 
$\omega_{x/y/z}(\vc k)=\lambda\sqrt{1+(\kappa/\kappa_c)\phi_{\vc k}}$ with    
$\phi_{\vc k}\!=\!\frac{2}{z} \sum_\gamma \cos(k_\gamma)$ has a finite gap 
$\lambda\sqrt{1-(\kappa/\kappa_c)}$. At $\kappa=\kappa_c$, the gap closes and, 
say, $T_z$ boson condenses to give a finite staggered magnetization 
$M_z=2\sqrt{6\rho(1-\rho)}$ at $\kappa >\kappa_c$, where 
$\rho=\frac{1}{2}(1-\tau^{-1})$ is the condensate density expressed 
via dimensionless parameter $\tau=\kappa/\kappa_c >1$. The $M$-length 
fluctuations, i.e., the amplitude ''Higgs'' mode, has a dispersion 
$\omega_z(\vc k)\simeq\lambda\sqrt{\tau^2+\phi_{\vc k}}$ with the gap 
$\Delta=\lambda\sqrt{\tau^2-1}$, while $T_{x/y}$ excitons become gapless 
Goldstone magnons with the energy 
$\omega_{x/y}(\vc k)\simeq \lambda\frac{\tau+1}{2}\sqrt{1+\phi_{\vc k}}\;$. 

We are ready to show our theory in action, by applying it to $d^4$ Mott 
insulator Ca$_2$RuO$_4$~\cite{Nak97} where a sizable value of the $LS$ product 
has indeed been observed~\cite{Miz01}. This fact implies the presence of 
unquenched spin-orbit coupling which is the basic input of our model. 

First, we compare the observed staggered moment 
$M\simeq 1.3\:\mu_\mathrm{B}$~\cite{Bra98} with our result 
$M=\sqrt{6(1-\tau^{-2})}\;\mu_\mathrm{B}$, and find $\tau\simeq 1.18$, 
i.e., this compound is rather close to the magnetic critical point. 
For spin-orbit coupling $\lambda (=\xi/2)\simeq 75$~meV~\cite{Miz01}, 
this translates into $\frac{4t^2}{U}\simeq53$~meV, a reasonable value for 
$t_{2g}$ systems with $t\sim 0.2$~eV and $U\sim 3-4$~eV. 

Second, using spin and orbital moments in the condensate 
$S=-L=\frac{1}{3\mu_{\rm B}}M$, we estimate their product $LS\simeq-0.2$ which 
is not too far from $-0.28\pm0.07$ observed~\cite{Miz01}. 

Third, we obtain from our theory the uniform magnetic susceptibility 
$\chi=\frac{12\:\mu_{\rm B}^2N_{\rm A}}{\lambda\:(1+\tau)}\simeq 
2.3\times 10^{-3}$~emu/mol, which is consistent with that of Ca$_2$RuO$_4$ 
($\sim 2.5\times 10^{-3}$~emu/mol~\cite{Nak97,Bra98}) above N\'{e}el 
temperature, where it is only weakly temperature dependent as expected 
for Van Vleck-type systems.

With the above numbers at hand (in fact, all extracted from the data), we 
predict the amplitude-mode gap $\Delta\sim 45$~meV, and the topmost 
energies $\sim 115$~meV for all three magnetic modes. We are not aware of 
inelastic magnetic data for Ca$_2$RuO$_4$ to date; resonant x-ray or neutron 
scattering experiments would provide a crucial test for the theory.  

We now turn to compounds with 90$^\circ$ $d$-$p$-$d$ bonding geometry, where 
effective interactions lead to remarkable features not present in perovskites. 
  
{\it Singlet-triplet model $H_{eff}(90^\circ)$}.-- This case is relevant to
delafossite {\it AB}O$_2$ or {\it A}$_2${\it B}O$_3$ structures where 
{\it B}O$_6$ octahedra share the edges and TM-ions form triangular or
honeycomb lattices (e.g., Li$_2$RuO$_3$).  
Using the same notations as above, we find  
\begin{eqnarray}
\label{eq:90h2}
h^{(c)}_2\!\!&=&\!3(\vc v_i\!\cdot\!\vc v_j - v_{iz}v_{jz}) 
-\frac{1}{3}(\vc u_i\!\cdot\!\vc u_j- u_{iz}u_{jz}), \\ 
&\equiv&\!\frac{2}{3}(T^\dagger_{ix}T_{jx}+T^\dagger_{iy}T_{jy})
-\frac{5}{6}(T_{ix}T_{jx}+T_{iy}T_{jy}) + H.c., \nonumber \\
\label{eq:90h3} 
h^{(c)}_3\!\!&=&\!\!\frac{1}{\sqrt{24}}(3\vc v_i\!\cdot\!\vc J_j \!+\!3v_{iz}J_{jz} 
-u_{ix}Q_{jx} \!+\! u_{iy}Q_{jy}) \!+\! (i\!\leftrightarrow\!j), \nonumber \\  
\label{eq:90h4}
h^{(c)}_4\!\!&=&\!-\frac{3}{4}d^\dagger_{ij}d_{ij} 
+\frac{1}{4}(J_{iz}J_{jz}\!+\!Q_{iz}Q_{jz})+\frac{1}{6}(n_in_{jz}\!+\!n_{iz}n_j)
\nonumber \\
&&-\frac{1}{12}n_in_j. \nonumber 
\end{eqnarray}
Again, $h^{(\gamma)}$ for $\gamma=a,b$ follow from $x,y,z$-cyclic permutations. 
While bond-dependent nature of $h^{(\gamma)}$ is expected for SOC models on
general grounds~\cite{Kha05}, it is surprising that interactions are of the 
same strength as in perovskites. This is in sharp contrast to 
$d^5(J\!=\!1/2)$ Kramers ions for which the leading exchange term vanishes 
and the form of the Hamiltonian is decided by smaller and competing effects 
of lattice distortions, Hund's coupling, higher-lying $e_g$ orbital, 
etc~\cite{Kha05,Che08,Jac09,Cha10}. Here, no cancellation of $d$-$p$-$d$ 
hoppings occurs and magnetic coupling is of the scale of $4t^2/U$ and hence 
strong, which is unusual for edge-shared oxides in general. This implies 
robustness of the physics discussed here against distortions, etc., an 
important point for the material design. 

\begin{figure}[tb]
\begin{center}
\includegraphics[width=8.2cm]{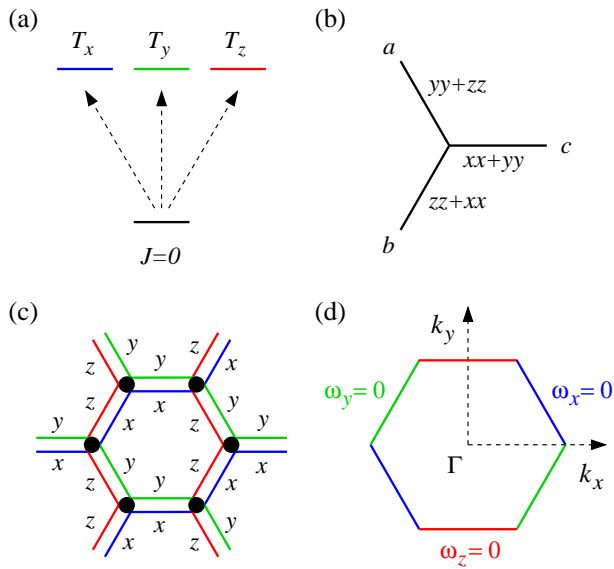}
\caption{(color online). 
Schematic of $T$-exciton dynamics in 90$^\circ$-bonding geometry. (a) Three 
types of $T$ excitations. (b) Three types of bonds on triangular or honeycomb 
lattices; $xx+yy$ indicates that only $T_x$ and $T_y$ bosons can move along 
$c$ bonds. (c) On a honeycomb lattice, each type of exciton forms its own 
zigzag chain, e.g., $z$ path for $T_z$. (d) In momentum space, each 
$T_\gamma$ boson softens and forms a quasicondensate at the respective 
edges of the Brillouin zone where $\omega_\gamma=0$~\cite{note_4}.    
}
\label{fig1}
\end{center}
\end{figure}

We start again with the quadratic part of $H_{eff}(90^\circ)$ which reads as  
\begin{eqnarray}
\label{eq:90}
H_2=\lambda\sum_in_i +\kappa\;\frac{1}{6}\sum_{ij} 
[(\vc T^\dagger_i \!\cdot\!\vc T_j \!\!&-&\!\! T^\dagger_{i\gamma}T_{j\gamma}) \\
-\frac{5}{8}(\vc T_i\!\cdot\!\vc T_j \!\!&-&\!\! T_{i\gamma}T_{j\gamma}+H.c.)]. 
\nonumber 
\end{eqnarray}
It is clear from Eqs.~\eqref{eq:90h2} and~\eqref{eq:90} 
[see also Fig.~\ref{fig1}(a,b)] that on each bond, two types of bosons are 
only active. Consequently, a given boson flavor $T_\gamma$ selects two types 
of bonds where it can move. For a triangular lattice, this is not crucial,  
though: Each flavor $T_\gamma$ forms its own square-type sublattice, so the 
vector field $\vc v$ eventually condenses into a conventional $120^\circ$ 
N\'{e}el ground state as soon as $\kappa>\kappa_c=\frac{4}{3}\lambda$. 
Details can be worked out along the lines of previous section; we focus 
instead on a honeycomb lattice below. 

On a honeycomb (''depleted'' triangular) lattice, each boson can move along 
a particular zigzag chain only, see Fig.~\ref{fig1}(c). Such a bond-and-flavor 
selective feature resembles that of the honeycomb Kitaev model~\cite{Kit06}, 
but we are dealing here with vector bosons, not spins, and this brings about
interesting new physics. 
  
In a paramagnetic phase, dispersion of the $T_z$ boson 
$\omega_z(\vc k)\equiv\omega_z(k_y)\simeq
\lambda\sqrt{1+(\kappa/\kappa_c)c_y}$ with $c_y=\cos(\frac{\sqrt 3}{2}k_y)$ 
is purely one-dimensional. As $\kappa\rightarrow\kappa_c=\frac{4}{3}\lambda$, 
a flat zero mode emerges at the Brillouin zone (BZ) edge 
$k_y=\pm\frac{2\pi}{\sqrt 3}$. Similarly, $T_x$ and $T_y$ boson dispersions 
collapse at the other edges corresponding to $x$ and $y$ zigzag directions, 
encircling thereby the BZ by the critical $\vc k$ lines [Fig.~\ref{fig1}(d)].

Thus, for $\kappa>\kappa_c$, strongly interacting, multicolor quasicondensate 
emerges. The fate of this critical system is decided by $h_{3,4}$ interactions 
in Eq.~\eqref{eq:90h2} which, in contrast to the previous section, become of 
paramount importance. In particular, the largest amplitude, bond-singlet 
density term $-d^\dagger_{ij}d_{ij}$ in $h_4$ deserves special attention. This 
interaction tends to bind the $\vc T$ bosons into singlet pairs, cooperating 
thereby with a pair-generation term $\vc T_i\!\cdot\!\vc T_j\propto d_{ij}$ in 
Eq.~\eqref{eq:90}. Put another way, its equivalent, i.e., biquadratic exchange 
$-(\vc J_i\!\cdot\!\vc J_j)^2$ is known to favor spin-singlet or spin-nematic 
states over magnetic order~\cite{Chu91,Dem02,Yip03}.       

We encounter here the rich and highly nontrivial problem which requires 
in-depth studies using the field-theory as well as numerical methods; this 
goes beyond the scope of the present work. We may, however, indicate potential 
instabilities and possible scenarios. 

The Hamiltonian~\eqref{eq:90h2} possesses a threefold symmetry (originating 
from $t_{2g}$ orbital degeneracy): $C_3$ rotation of the lattice and permutation 
among $T_x,T_y,T_z$ flavors. This discrete symmetry can be broken at finite
temperature. One may expect at least three distinct ground states as a function 
of $\kappa$: a trivial paramagnet below $\kappa_{c1}\lesssim\frac{4}{3}\lambda$; 
a long-range magnetic order when boson density becomes large at 
$\kappa_{c2}>\frac{4}{3}\lambda$; and an intermediate phase at 
$\kappa_{c1} <\kappa <\kappa_{c2}$ hosting the spin singlet GS. The most 
intriguing option for the latter is a spin-superfluid state, often discussed 
in the context of spin-one bosons~\cite{Ess09} and bilinear-biquadratic spin 
models~\cite{Ser11}. Here, this state is supported by a flavor-symmetric  
attraction $-d^\dagger_{ij}d_{ij}$ between $T_\gamma$, selecting a global spin 
singlet of $A_{1g}$ symmetry where all three flavors form pairs and condense, 
but there is a single-particle gap. Another possibility, favored by the 
bond-directional nature of hoppings and interactions in Eq.~\eqref{eq:90h2}, 
is a nematic order, i.e. spontaneous selection of a particular zigzag out of 
three $x,y,z$ directions (assisted in real systems by electron-lattice 
coupling). Once zigzag chain is formed, it can dimerize due to biquadratic 
interactions~\cite{Chu91,Yip03,Liu12}. In other words, boson pairs condense 
into a valence-bond-solid pattern, followed by a suppression of Van Vleck 
susceptibility. Future studies are necessary to clarify the phase behavior 
of the model near the magnetic critical point. 

In the recent past, some unusual properties of ruthenate compounds have been
reported, including the formation of one-dimensional, spin-gapped 
chains in Tl$_2$Ru$_2$O$_7$~\cite{Lee06}, and singlet dimers in 
La$_4$Ru$_2$O$_{10}$~\cite{Khal02,Hua06}. Of particular interest is 
a honeycomb lattice La$_2$RuO$_3$~\cite{Miu07} which forms dimerized zigzag 
chains. This observation has been discussed in terms of orbital 
ordering~\cite{Jac08}; the present model based on spin-orbit coupling may 
provide an alternative way. Indeed, $z$-type zigzag chain dimerized by 
biquadratic exchange (and supported by electron-lattice coupling) would give 
the same pattern as observed~\cite{Miu07}. 
Future experiments, in particular a direct measurement of the $LS$ product, 
should tell whether an unquenched $L$ moment (the key ingredient of our 
model) is present in these compounds, in order to put the above ideas 
on a more solid ground. 

On theory side, apart from low-energy properties of the model itself, 
important questions are related to doping of Van Vleck-type $d^4$ insulators. 
A doped electron, i.e. $J=1/2$ fermion moving on a background of 
singlet-triplet $d^4$ lattice, should have a large impact on magnetism and 
vice versa. Unconventional pairing via the exchange of $T$ excitons 
also deserves attention, in particular, on triangular and honeycomb 
lattices (where unusual pairing symmetries have been suggested for  
$d^5$ systems with strong SOC~\cite{Kha04,Kha05,Hya12,You12}). 
We recall that energy scales involved in $d^4$ systems are large even for 
$90^\circ$ $d$-$p$-$d$ bonding, so all the ordering phenomena are expected 
at higher temperatures than in $d^5$ compounds like triangular 
lattice Na$_x$CoO$_2$ or honeycomb Na$_2$IrO$_3$. 

To conclude, unconventional magnetism emerging from exciton condensation, 
rather than from orientation of the preexisting local moments, can be realized 
in Mott insulators of Van Vleck-type ions with a nonmagnetic ground state. 
We derived effective models describing the magnetic condensate and its 
elementary excitations on various lattices. Of particular interest is the
emergence of quasi-one-dimensional condensate of magnetic excitons on 
a honeycomb lattice. We discussed implications of the theory for 
candidate Van Vleck-type Mott insulators.  

We thank J. Chaloupka, B.J. Kim, and A. Schnyder for discussions.


\begin{thebibliography}{99}
\bibitem{note_uv} The $(\vc u,\vc v)$ representation is often more convenient, 
in particular for semiclassical analysis. Physically, $\vc v$ takes care of 
a magnetic dipole carried by the $\vc T$ exciton (and hence enters in 
$\vc S, \vc L$, and $\vc M$), while $\vc u$ stands for the quadrupolar 
component of $\vc T$.
\bibitem{note_J} In the $(\vc u,\vc v)$ representation, 
$\vc J= 2(\vc u\times\vc v)$. We will see that $\vc v$ condenses at the 
critical point but $\vc u$ remains fluctuating.
\bibitem{note_ni} We neglect here a small correction to singlet-triplet 
splitting $\lambda$ arising from single-ion energy shifts; for a cubic 
lattice, the renormalization is $\lambda \rightarrow \lambda-\frac{2t^2}{3U}$,
and even smaller for the low-coordination square and honeycomb lattices.
\bibitem{note_Q} Quadrupoles $Q$ arise from the mapping of composite 
spin-orbital terms $S^\alpha L^\beta$ in $H$, e.g., 
$S^yL^z=\!\sqrt\frac{2}{3}\;u_x +\frac{1}{2}Q_x$. Note that $\vc u$ enters 
here, revealing its quadrupole nature.
\bibitem{note_av} The only effect of small ''compass-like'' 
$v_{i\gamma}v_{j\gamma}$ terms here is to select an easy magnetic axis 
and open (order-by-disorder) magnon gap that can be routinely 
worked out~\cite{Kha01}.
\bibitem{note_4} We may reverse the sign of $\kappa$ in Eq.~\eqref{eq:90} by 
virtue of the four-sublattice transformation~\cite{Kha05}. Then, zero-energy 
lines would go through the $\Gamma$ point, e.g., $\omega_z(\vc k)=0$ at 
$k_y=0$.
\end{thebibliography}
\end{document}